\documentstyle[12pt]{article}
\begin{document}
\begin{center}
{\large \bf ABS Methods and ABSPACK for Linear Systems and Optimization, a Review }
\vskip20mm
 Emilio Spedicato, Elena Bodon, Antonino Del Popolo\\
 Department of Mathematics,  University of Bergamo, Bergamo \\
\vskip6mm
Nezam Mahdavi-Amiri,  \\
Department of Mathematics, Sharif University, Tehran\\
\end{center}
\vskip20mm
Address of person to whom send the proofs: Emilio Spedicato, Department of Mathematics, University of Bergamo, via dei Caniana 2, 24128 Bergamo, emilio@unibg.it
\vskip20mm
Running title: A review of ABS methods

\newpage
\begin{abstract}
ABS methods are a large class of methods, based upon the Egervary rank reducing algebraic process, first introduced in 1984 by Abaffy, Broyden and Spedicato for solving linear algebraic systems, and later extended to nonlinear algebraic equations, to optimization problems and other fields; software based upon ABS methods is now under development.
Current ABS literature consists of about 400 papers. ABS methods provide a unification of several classes of classical algorithms and more efficient new solvers for a number of problems. In this paper we review ABS methods for linear systems and optimization, from both the point of view of theory and the numerical performance of ABSPACK.
\end{abstract}
\noindent
\section{ Introduction}

ABS algorithms were  introduced by Abaffy, Broyden and Spedicato (1984), 
 to solve linear equations first in the form of the
 {\it basic ABS class}, later generalized 
 as the  {\it scaled ABS class}. They were then 
 applied  to linear least squares, nonlinear equations 
and optimization problems, see e.g. the monographs by
Abaffy and Spedicato (1989) and Feng et al. (1998), or the
  bibliography by Spedicato  et al.
  (2000) listing  350 ABS papers. In this paper we  
 review some of the main results obtained in the field of
 ABS methods and we provide some results on the 
 performance of  ABSPACK, a
 FORTRAN package  based on ABS methods presently under development.

 The main results obtained in almost twenty years of research in the ABS field

 can be summarized as follows:
 \begin{itemize}
 \item  ABS methods provide a unification of the field of finitely terminating
 methods for linear systems and for feasible direction
 methods for linearly constrained optimization, see sections 2 and 6; due to
 the several alternative formulations of their algebra they  lead to
  different computational implementations of the algorithms, each one
 with a special feature that may  be of advantage
 \item ABS methods  provide some new methods that are better than classical
 ones under several respects. For instance the implicit LX method requires
 the same number of multiplications as Gaussian elimination (which is optimal
 under very general conditions) but requires less memory and does not  need pivoting
 in general; moreover its application to the simplex method has not only a
 lower memory requirement,  but is cheaper in multiplications up to one order
 with respect to e.g. the Forrest-Goldfarb implementation of the simplex
 method via the LU factorization, see section 5. Also there are ABS methods for nonlinear
 problems that are  better than  Newton method in terms of
 convergence speed (some ABS algorithms keep quadratic rate of convergence even if the
 Jacobian is singular at the solution) or in terms  of required information (about $n^2/2$ Jacobian
 components against the full $n^2$)
 \item on some classes of significant problems ABSPACK codes have a better performance
 in both accuracy and speed that the codes in LAPACK (usually considered
 the best package now on the market); see section 7.
 \item ABS methods have allowed to solve open problems in the literature
 (e.g. the explicit determination  of Quasi-Newton updates for the
 sparse symmetric case, see Spedicato and Zhao (1993))
 \item for linear Diophantine equations ABS methods have led to a new
 solution method, that also provides the first  generalization of
 the classical existence theorem  of Euclides, Diophantus and Euler from a
 single $n$-dimensional equation to a general $m$-equations system in
 $n$ variables, see section 4.
 \end{itemize}

\section{The scaled ABS class: definition and main properties}

 Let us consider the following
 determined or underdetermined linear system, where
$rank(A)$ is arbitrary and $ A^T = (a_1,\ldots,a_m) $

$$
Ax = b \ \ \ \ \  x \in {R}^n ,\ \  b \in {R}^m , \ \ m \leq n \eqno {(1)}
$$
or
$$
a_i^Tx - b_i = 0, \ \ \  i = 1,\ldots,m  \eqno {(2)}
$$

The above system can be solved by the following {\it scaled ABS class}
of  algorithms, which can be shown to provide a complete realization
of the general class of methods implementing the {\it Petrov-Galerkin}
criterion  (stating that the residual vector at the $i$-th iteration
is orthogonal to a 
given arbitrary set of $i-1$ linearly independent vectors).

\vskip2mm
\noindent {\bf The scaled ABS algorithm}
\vskip2mm
\begin{description}
\item[(A)] Give $x_1 \in {R}^n$  arbitrary,\ $H_1 \in {R}^{n,n}$ 
nonsingular arbitrary,  $v_1 \in R^m$  arbitrary nonzero.
Set $i = 1$.
\item[(B)] Compute the residual $r_i = Ax_i-b$. If $r_i = 0$ stop ($x_i$
solves the problem.) Otherwise compute
$s_i = H_iA^Tv_i$. If $s_i \neq 0$, then  go  to (C).  If $s_i = 0$
and $\tau = v_i^Tr_i = 0$, then set $x_{i+1} = x_{i} ,\  H_{i+1} = H_{i}$
(the $i$-th equation is redundant) 
and go to (F). Otherwise stop (the system has no solution).
\item[(C)] Compute the search vector $p_i$ by
$$
p_i = H_i^Tz_i \eqno{(3)}
$$
where $z_i \in  {R}^n$ is arbitrary save for the condition
$$
v_i^TAH_i^Tz_i \neq 0 \eqno{(4)}
$$
\item[(D)] Update the estimate of the solution by
$$
x_{i+1} = x_{i} - \alpha_ip_i, \ \ \ \alpha_i = v_i^Tr_i/v_i^TAp_i.  \eqno{(5)}
$$
\item[(E)] Update the matrix $H_i$ by
$$
H_{i+1} = H_i - H_iA^Tv_iw_i^TH_i/w_i^TH_iA^Tv_i \eqno{(6)}
$$
where $w_i \in {R}^n$ is arbitrary save for the condition
$$
w_i^TH_iA^Tv_i \neq 0.   \eqno{(7)}
$$
\item[(F)] If $i = m$,  stop  ($x_{m+1}$ solves the system). Otherwise 
give  $v_{i+1} \in R^m $  arbitrary linearly
independent from $v_1, \ldots, v_i$. Increment
$i$ by one and go to ({\bf B}).
\end{description}
Matrices $H_i$, which  are generalizations of  
(oblique) projection matrices, have been named
 {\it Abaffians} at the First International Conference on ABS methods
 (Luoyang, China, 1991). However we must recall that these matrices
  were used in  several little known papers by Egervary, predating
 the ABS algorithms. There are  alternative formulations of the
 scaled ABS algorithms, e.g. using
 vectors instead of the  square matrix $H_i$, with possible
 advantages in storage and number of operations. 
For details see Abaffy and Spedicato (1989) and on how these formulations
can be used to imbed in an ABS approach even iterative methods (including the
Kaczmarz, Gauss-Seidel, ORTHOMIN, ORTHODIR etc. methods), see Spedicato
and Li (2000).

 The  choice of the parameters $H_1, \ v_i, \ ,z_i, \  w_i$ determines
 particular methods.
The {\it basic ABS class} is obtained by taking $v_i = e_i, $ as the
$i$-th unit vector in $R^m$. 

We recall some  properties of  the scaled ABS class,
assuming that $A$ has full rank.
\begin{itemize}
\item Define  $V_i = (v_1,\ldots,v_i)$ and
$W_i = (w_1,\ldots,w_i)$. Then
$$
H_{i+1}A^TV_i = 0, \ \  H_{i+1}^TW_i = 0  \eqno{(8)}
$$.
\item The vectors $H_iA^Tv_i, \  H_i^Tw_i $ are zero 
if and only if $a_i, \ w_i$ are respectively linearly dependent from
$a_1,\ldots,a_{i-1}, \ w_1, \ldots, w_{i-1}$. 
\item Define  $P_i = (p_1,\ldots,p_i) $ and $A_i = (a_1, \ldots,a_i)$.
Then the  implicit factorization 
$V_i^TA_i^TP_i = L_i$ holds, where $L_i$ is nonsingular lower triangular.
Hence, if $m=n$, one obtains a semiexplicit 
factorization of the inverse, with $P = P_n, \ V = V_n, \ L = L_n$
$$
A^{-1} = PL^{-1}V^T.  \eqno{(9)}
$$
For several choices of  $V$ the matrix $L$ is diagonal, hence
formula (8) gives a fully explicit factorization of the inverse as a
byproduct of the ABS solution of a linear system.

\item The general solution of system (1) can be written as follows,
with $q \in R^n$ arbitrary
$$
x = x_{m+1} + H_{m+1}^Tq    \eqno{(10)}
$$
\item The Abaffian can be written in terms of the parameter matrices
as
$$
H_{i+1} = H_1 - H_1A^TV_i(W_i^TH_1A^TV_i)^{-1}W_i^TH_1. \eqno{(11)}
$$
\end{itemize}
Letting $V = V_m, \ W = W_m$, one can show that the parameter matrices $H_1, \ 
V, \ W$ are admissible (i.e.  condition (7) is satisfied) iff
the matrix $Q = V^TAH_1^TW$ is strongly nonsingular (i.e. it 
is LU factorizable).
This condition can  be satisfied by  exchanges
of the columns of $V$ or $W$.
If $Q$ is strongly nonsingular and we take, as is 
done in all algorithms so far considered, $z_i = w_i$, then condition
(4) is also satisfied. Analysis of the conditions under which $Q$ is not
strongly nonsingular leads, when dealing with Krylov space 
methods in their ABS formulation, to a characterization of the topology
of the starting points that can produce a breakdown (either  a division by
zero or  a vanishing search direction) and to several ways of curing it,
 including those considered in the literature.

Three subclasses of the scaled ABS class and  particular algorithms are
now recalled.

\begin {description}
\item[(a)] The {\it conjugate direction subclass}.
This class is obtained by setting $v_i = p_i$. It 
is well defined under the condition (sufficient but not necessary)
that $A$ is symmetric and positive definite. It contains 
the ABS versions of
the  Choleski,
the Hestenes-Stiefel and the Lanczos algorithms. This class generates
all possible algorithms whose search directions are $A$-conjugate. 
If $x_1=0$, the vector
$x_{i+1}$ minimizes the energy ($A$-weighted Euclidean)
norm of the error over  $ Span(p_1,\ldots,p_i)$ and the
solution is approached monotonically from below in the energy norm.

\item[(b)] The {\it orthogonally scaled subclass}.
This class is obtained by setting $v_i = Ap_i$. It 
is well defined if $A$ has full column rank and remains 
well defined even if $m$ is greater than
$n$. It contains the ABS formulation of the QR algorithm
(the so called {\it implicit QR algorithm}),  the GMRES and 
the conjugate residual algorithms. The scaling vectors are orthogonal and
the search vectors are $A^TA$-conjugate. If $x_1=0$, 
the vector $x_{i+1}$ minimizes the
Euclidean norm of the residual over $ Span(p_1, \ldots,p_i)$ and
the solution is monotonically approached from below in the residual norm.
It can be shown that
the methods in this class can be applied to overdetermined systems 
of $ m > n$ equations, where in $n$ steps they  obtain
the solution in the least squares sense. 
\item[(c)] The {\it optimally stable subclass}. This class is obtained by setting
$v_i = (A^{-1})^Tp_i$, the inverse disappearing in the actual recursions. The
search vectors in this class are orthogonal.
If $x_1=0$, then
the vector $x_{i+1}$ is the vector of least Euclidean norm over
$ Span(p_1, \ldots,p_i)$ 
and the solution is approached
monotonically from below in the  Euclidean norm. The methods of Gram-Schmidt and
of Craig belong to this subclass. The methods in this class have minimum
error growth in the approximation to the solution according to a criterion
by Broyden.
\end{description}

\section{   Special algorithms in the basic ABS class}
In this section we consider in more detail three specially important
algorithms, that belong to  the basic ABS class (i.e. $V=I$).
\begin{description}
\item[(A)] 
The {\it Huang algorithm} is obtained by the  choices $H_1 = I$, $z_i =
w_i = a_i$. 
A mathematically equivalent, but numerically more stable, formulation
is the so called {\it modified Huang algorithm} 
($p_i = H_i(H_ia_i)$ and
$H_{i+1} = H_i - p_ip_i^T/p_i^Tp_i$). 
Huang algorithm generates
 search vectors that are orthogonal and identical with those
obtained by the  Gram-Schmidt  procedure applied
to the rows of $A$. 
 If $x_1=0$, then  $x_{i+1}$ is the solution
with least Euclidean norm of the first $i$ equations. The solution $x^+$ with
least Euclidean norm of the whole system is approached monotonically and from below
by the sequence $x_i$.  For arbitrary starting point the Huang algorithm generates
that solution  which is closest in Euclidean norm to  the initial point.
\item[(B)] 
The {\it implicit LU algorithm} is given by
the choices $H_1 = I,\  z_i = w_i = e_i$.
It is well defined iff $A$ is regular (i.e. all 
principal submatrices are nonsingular). Otherwise column
pivoting  has to be performed (or, if $m = n$, equation pivoting).
The Abaffian  has the following structure, with
$K_i \in R^{n-i,i}$
$$H_{i+1} = \left
	  [
	  \begin{array}{cc}
	  0      & 0 \\
	  \cdots & \cdots \\
	  0      & 0 \\
	  K_i    & I_{n-i}
	  \end{array}
	  \right ].     \eqno{(12)}
$$
implying that the matrix $P_i$ is
unit upper triangular, so that the implicit factorization $A = LP^{-1}$ is of
the LU type, with units on the diagonal.
 The algorithm requires 
 for $m  =  n$,  
\  $n^3/3$
multiplications plus lower order terms,  the same cost of
   classical LU factorization or Gaussian
elimination. 
Storage requirement  for $K_i$
 requires at most  $n^2/4$ positions, i.e.
 half the storage needed by Gaussian elimination and a fourth that
  needed by the LU factorization algorithm
(assuming that $A$ is not overwritten). Hence the implicit LU algorithm
has same arithmetic cost but uses less memory than the most efficient
classical methods.
\item[(C)] 
The {\it implicit LX algorithm}, see Spedicato, Xia and Zhang (1997), 
is defined by the choices
$H_1 = I,\ z_i = w_i = e_{k_i}$,
where $k_i$ is an integer, $1 \leq k_i \leq n$, such that
$e_{k_i}^TH_ia_i \neq 0. $ 
By a general property of the ABS
class for $A$ with full rank there is at least
one index $k_i$ such that $e_{k_i}^TH_ia_i \neq 0. $ For stability reasons
we select $k_i$ such
that $\mid e_{k_i}^TH_ia_i \mid$ is maximized. This algorithm
has the same overhead and memory requirement as the implicit LU algorithm,
but does not require  pivoting. Its computational performance is also
superior and generally better than the performance of  the classical LU
factorization algorithm with row pivoting, as available for instance in
 LAPACK or MATLAB, see Mirnia (1996). Therefore this algorithm can be
considered as {\it the most efficient 
general purpose linear solver not of the Strassen type}.

\end{description}

\section{ Solution of linear Diophantine equations}

One of the main results  in the ABS field  has been the derivation of   
ABS methods for linear
Diophantine equations. The ABS algorithm determines 
if the Diophantine system has an
integer solution, computes a particular solution and provides a
representation of  all integer solutions. It is a generalization of
  a method proposed by Egervary (1955) for the 
 particular case of a homogeneous system.

Let ${Z}$ be the set of all integers and consider the Diophantine linear
system of equations
$$
Ax=b, \quad  \quad x\in {Z}^n,             \
A\in {Z}^{m\times n}, \ b\in {Z}^m, \ m\leq n.\eqno{(13)}
$$

While  thousands of  papers have been
written concerning nonlinear, usually polynomial, Diophantine equations
in few variables, the general linear system has attracted   much
less attention. The single  linear equation in $n$ variables was first
solved by Bertrand and Betti (1850). Egervary  was probably
the first author dealing with  
a system (albeit only the homogeneous one).  Several methods for 
the nonhomogeneous system have recently been proposed based mainly on
reduction to canonical forms.

We  recall some results from number theory.
Let $a$ and $b$ be integers. If there is integer $\gamma$ 
so that $b= \gamma a$ then
we say that $a$ divides $b$ and write $a|b$, otherwise 
we write $a \!\!\not|\ b$.
If $a_1,\dots,a_n$ are integers, not all being zero, then the greatest common
divisor ($gcd$) of these numbers is the greatest positive integer $\delta$ which
divides all
$a_i$, $i=1,\dots,n$ and we write $ \delta =gcd(a_1,\dots,a_n)$. We
note that
$\delta \geq 1$ and that 
$\delta$ can be written as an integer linear combination of
the $a_i$, i.e. $\delta = z^Ta$ for some $z\in R^n$.
One can show that $\delta$ is the least positive integer for which
the equation $a_1x_1+\dots+a_nx_n= \delta $ has an integer solution. Now
$\delta$ plays a main role in the following
\vskip3mm
	 
\noindent
{{\bf  Fundamental Theorem of the  
Linear Diophantine Equation}}
\vskip2mm
{\it
\noindent Let $a_1,\dots,a_n$ and $b$ be integer numbers. Then the
 Diophantine linear equation $a_1x_1+\dots+a_nx_n=b$
has  integer solutions if and only
if $gcd(a_1,\dots,a_n) |\ b$. 
In such a case if $n>1$ then
there is an infinite number of integer solutions. }

\vspace{0.5cm}

\noindent In order to find the general integer solution of the Diophantine
equation $a_1x_1+\dots+a_nx_n=b$, the main step is to solve
$a_1x_1+\dots+a_nx_n= \delta$,
where $\delta=gcd(a_1,\dots,a_n)$, for a special integer solution. 
There exist several
algorithms for this problem.
The basic step is the computation of $\delta$ and $z$,
often done using  the  algorithm of
Rosser (1941), which  
 avoids a too rapid growth of the intermediate integers, and which
 terminates in polynomial time, as shown
by Schrijver (1986). The scaled ABS algorithm can be applied to Diophantine  equations via a special
choice of its parameters, originating from the following considerations
and Theorems.

Suppose $x_i$ is an integer vector.
Since $x_{i+1}=x_i-\alpha_ip_i$, then $x_{i+1}$ is integer if
$\alpha_i$ and $p_i$ are integers. If $v_i^TAp_i|(v_i^Tr_i)$,
 then $\alpha_i$ is an integer. If $H_i$
and $z_i$ are respectively an integer
matrix and an integer vector, 
then $p_i=H_i^Tz_i$ is also an integer vector. Assume
$H_i$ is an integer matrix. From (6), if
$v_i^TAH_i^Tw_i$ divides all the components of $H_iA^Tv_i$, then $H_{i+1}$
is an integer matrix. 

Conditions for the existence of an integer solution and determination of all 
integer  solutions of the Diophantine system are given in the following
theorems,  generalizing the Fundamental Theorem, see 
Esmaeili, Mahdavi-Amiri and Spedicato
 (2001a).
\vskip3mm
{\bf Theorem 1} \
 {\it Let $A$ be full rank and suppose that the Diophantine
system (13) is integer solvable. Consider the  Abaffians generated
by the scaled ABS algorithm with the  parameter choices:
  $H_1$ is unimodular (i.e. both  $H_1$ and $H_1^{-1}$ are integer
  matrices);
  for $i = 1, \ldots, m$, \  $w_i$ is such that
$w_i^TH_iA^Tv_i = \delta_i,  \ \ \delta_i = gcd(H_iA^Tv_i)$.
Then the following properties are true:
\begin{description}
\item[(a)] the  Abaffians generated by the algorithm are
well-defined and are integer matrices
\item[(b)] if $y$ is a special integer solution of the first $i$
equations, then any integer solution $x$ of such equations can be
written as $x = y + H_{i+1}^Tq$ for some integer vector
$q$.
\end{description}    }
\vskip5mm  
{\bf Theorem 2} \
 {\it 
Let $A$ be full rank and consider the sequence of
matrices $H_i$ generated by the scaled ABS algorithm with parameter
choices as in Theorem 1. Let  $x_1$ in the scaled
ABS algorithm be an arbitrary integer vector and let $z_i$ be  such that
$z_i^TH_iA^Tv_i = gcd(H_iA^Tv_i)$.
Then system (13)         has integer solutions iff 
$gcd(H_iA^Tv_i)$ divides $v_i^Tr_i $ for $i = 1, \ldots, m$.  }
\vskip3mm
\noindent From the above theorems we obtain the following 
scaled ABS algorithm for Diophantine equations.
\vskip3mm
{\bf The  ABS Algorithm for  Diophantine Linear Equations}
\begin{enumerate}
\item[(1)] Choose $x_1\in{Z}^n$, arbitrary,  $H_1\in {Z}^{n\times n}$,
      arbitrary  unimodular. Let $i=1$.
\item[(2)] Compute $\tau_i=v_i^Tr_i$ and $s_i=H_iA^Tv_i$.
\item[(3)] {{\it If}} ($s_i=0$ and $\tau_i=0$) {{\it then}} let $x_{i+1}=x_i$,
      $H_{i+1}=H_i$, $r_{i+1}=r_i$ and {{\it go to}} step (5) (the $i$th
      equation is redundant).
      {{\it If}} ($s_i=0$ and $\tau_i\not =0$) {{\it then}} Stop (the
      $i$th equation and hence the system is incompatible).
\item[(4)] $\{s_i\not =0\}$ Compute  $\delta_i=gcd(s_i)$
      and $p_i=H_i^Tz_i$, where
      $z_i\in {Z}^n$ is an arbitrary integer vector satisfying
      $z_i^Ts_i=\delta_i$. {{\it If}} \ \
      $\delta_i\!\!\not| \tau_i$ {{\it then}} Stop 
      (the system is integerly inconsistent),
     {{\it else}}
      Compute
      $
      \alpha_i=\tau_i/\delta_i,
      $
      let
      $
      x_{i+1}=x_i-\alpha_ip_i
      $
 and update $H_i$  by
      $
      H_{i+1}=H_i-\frac{H_iA^Tv_iw_i^TH_i}{w_i^TH_iA^Tv_i}
      $
      where $w_i\in {R}^n$ is an arbitrary integer vector satisfying
       $w_i^Ts_i=\delta_i$.
\item[(5)] {{\it If}} $i=m$ {{\it then}} Stop ($x_{m+1}$ is a solution)
{{\it else}} let $i=i+1$ and {{\it go to}} step (2).
\end{enumerate}

\vspace{0.5cm}

\noindent It follows from Theorem 1 that 
if there exists a solution for the system (13), then
$x=x_{m+1}+H_{m+1}^Tq$, with arbitrary $q\in{Z}^n$, provides all
solutions of (13).  

Egervary's algorithm  for 
homogeneous Diophantine systems corresponds to the choices 
$H_1=I,\ x_1=0$ and
$w_i=z_i$, for all $i$. Egervary claimed, without  proof, that any set of
$n-m$ linearly independent rows of $H_{m+1}$ form an integer basis for the
general solution of the system. Esmaeili et al. (2001a) have shown by a counterexample that
Egervary's claim is not true in general; we have also provided an
analysis of conditions under 
which $m$ rows in $H_{m+1}$ can be eliminated.

One can show that  by special choices of the scaling parameters and by
an inessential modification of the update of the Abaffian it  is not
necessary to solve in general $2m$ single $n$-dimensional Diophantine 
equations (to determine $z_i, \ w_i$), but just one single Diophantine
equation, at the last step.

The ABS algorithm for linear Diophantine systems has been extended also
to systems of $m$ linear inequalities ($m \le n$), where it provides an
almost explicit representation of all solutions, see Esmaeili, Mahdavi-Amiri
and Spedicato (2001b). The used technique allows, in the case of $m \le n$
real inequalities to find a fully explicit representation of all solutions, see Esmaeili, Mahdavi-Amiri and Spedicato (2000), thereby obtaining a generalization of formula (10).  Finally, the ABS approach provides a new way of computing, given an integer matrix, its integer null space and, under some conditions, an integer basis of it, see Esmaeili, Mahdavi-Amiri and Spedicato (2001c).

\section{Reformulation of the simplex method via the implicit LX method}

\noindent The implicit LX algorithm has a natural application in reformulating the
simplex method for the LP problem in standard form:
$$ min \ c^Tx$$ subject to
$$Ax = b, \ \ x \ge 0.$$

The suitability of the implicit LX algorithm derives from the fact that
the algorithm, when started with $x_1=0$, generates basic type vectors
  $x_{i+1}$, which are vertices of the polytope defined by the
  constraints of the LP problem if the components not identically zero
  are nonnegative.

The basic step of the simplex method is moving from one vertex to another
according to certain rules and reducing at each step the value of
$c^Tx$.
The direction of  movement  is obtained by solving a linear system,
whose coefficient matrix  $A_B$, the {\it  basic matrix}, is defined by
 $m$ linearly independent columns of   $A$, called the  {\it basic columns}.
This system is usually solved via the LU factorization method or also via
the QR factorization method. 
The new vertex is associated with a new basic matrix
  $A_B'$, obtained by substituting one of the columns of 
 $A_B$ with a column of the matrix
$A_N$ that comprises the columns of $A$ not belonging to
 $A_B$. One has then to solve a new system where just one column has been
 changed. If the LU factorization method is used, then the most efficient
 way to recompute the modified factors is the steepest edge method
 of Forrest and Goldfarb (1992), that requires
 $m^2$ multiplications. This implementation of the simplex method
has the following storage requirement:
$m^2$ for the LU factors and   $mn$ for the matrix
 $A$, that has to be kept to provide the columns needed for the
exchanges.

The reformulation of the simplex method via the implicit LX method, see
Zhang and Xia (1995) and Spedicato,  Xia and Zhang (1997),
 exploits the fact that in the implicit LX
 method the exchange of the  $j$-th column in $A_B$
with the $k$-th column in $A_N$ corresponds to exchanging previously chosen
parameters
 $z_j = w_j = e_{j_B}$ 
with new parameters  $z_k = w_k =
e_{k_B}$. in terms of the implied modification of the Abaffian this operation is a
special case of a general rank-one modification of the parameter matrix
$W=(w_{k_1}, \ldots,w_{k_m})$. The modified Abaffian can be efficiently evaluated
using a general formula of 
 Zhang (1995), without explicit use of the $k$-th column of
 $A_N$. Moreover all information needed to build the search
 direction (the polytope edge) and to implement the (implicit) column
 exchange is contained in the Abaffian matrix. Thus
 there is no need to keep the matrix $A$. Hence storage requirement is
 only that needed in the construction and keeping of the matrix
  $H_{m+1}$, i.e. respectively
 $n^2/4$ and $n(n-m)$. For   $m$ close to  
 $n$, such storage is about 8 times less than in the Forrest and Goldfarb implementation
 of the LU method. For small $m$ storage is higher, but there is an
 alternative formulation of the implicit LX method that has a similar
 storage, see Spedicato and Xia (1999).

We now give the main formulas for the simplex method in the classical and
in the ABS formulation. The column in
 $A_N$  that  substitutes a column in
 $A_B$ is usually taken as the column with least relative cost.
In the ABS approach this corresponds to minimize with respect to
  $i \in N_m$ the scalar
$\eta_i = c^TH^Te_i$.
Let $N^\ast$ be the index so chosen. The column in
 $A_B$ to be exchanged is often chosen with the criterion of 
the least edge displacement that keeps the basic variables nonnegative.
Defining
$\omega_i = x^Te_i/e_i^TH^Te_{N^\ast} $
with $x$ the current vertex,  the above criterion is equivalent
to minimize
 $\omega_i$  with respect to the set of indices $i \in B_m$ such that
$$
e_i^TH^Te_{N^\ast} > 0 
\eqno{(14)}
$$
Notice that $H^Te_{N^\ast} \neq 0$ and that an index $i$ such that (14) be
satisfied always exists, unless  $x$ is a solution.

The update of the Abaffian after the exchange of the obtained unit vectors
is given by the following simple rank-one correction
$$
H' = H - (He_{B^\ast} - e_{B^\ast})e_{N^\ast}^{\ T}H/e_{N^\ast}^{\ T}He_{B^\ast}
\eqno{(15)}
$$
The displacement vector $d$, classically obtained by solving system
  $A_Bd = - Ae_{N^\ast}$, is obtained at no cost by
$d = H_{m+1}^{\ T}e_{N^\ast}$. The relative cost vector, classically given by
$r = c - A^TA_B^{-1}c_B$,
with $c_B$ comprising the components of $c$ with indices corresponding to
those of the basic columns, is given by formula
 $r = H_{m+1}c$.

It is easily seen that update (15) requires no more than
 $m(n-m)$ multiplications.  Cost is highest for  $m = n/2$ 
 and gets smaller as  $m$ gets smaller or closer to
  $n$. 
In the  method of  Forrest and
 Goldfarb  $m^2$ multiplications are needed to update the LU
 factors and again $m^2$ to solve the system. Hence the ABS approach via
 formula (15) is faster for $m>n/3$. For $m<n/3$ similar costs are obtained
 using the alternative formulation of the implicit LX algorithm described
 in Spedicato and Xia (1999). For $m$ very close to $n$ the advantage of the
 ABS formulation is essentially of one order and such that 
 no need is seen to develop formulas for the sparse case.

There is an ABS generalization of the simplex method based upon a modification
of the Huang algorithm which is started by a certain singular matrix, see
Zhang (1997). In such generalization the solution is approached via a 
sequence of points lying on the faces of the polytope. If one of such points
happens to be a vertex, then all successive iterates are vertices and
 the standard simplex method is reobtained.

\section{ABS unification of feasible direction methods for linearly constrained
minimization}

\noindent ABS methods allow a unification of feasible direction methods for linearly
constrained minimization, of which the LP problem is a special case. Let
us first consider the problem with only equality constraints
$$min \ f(x), \ \ x \in R^n$$
subject to
$$ Ax = b, \ \ A \in R^{m,n}, \ \ m \le n,\ \ rank(A) = m.$$

Let $x_1$ be a feasible initial point. If we consider an iteration of the
form
$x_{i+1} = x_i - \alpha_id_i$,  the sequence $x_i$ consists of feasible
points iff
$$
Ad_i = 0  \eqno{(16)}
$$
The general solution of (16) can be written using the ABS formula
(10)
$$
d_i = H_{m+1}^Tq  \eqno{(17)}
$$
In (17) matrix $H_{m+1}$ depends on  parameters 
$H_1$, $W$ and $V$ and $q \in {R^n}$ can also be seen
as a parameter. Hence the general iteration generating feasible points is
$$
x_{i+1} = x_i - \alpha_iH_{m+1}^Tq.  \eqno(18)
$$
The search vector is a descent direction if 
$d^T\nabla{f(x)} = q^TH_{m+1}
\nabla f(x) > 0$. This condition can always be satisfied by a choice of 
 $q$ unless
$H_{m+1}\nabla f(x) = 0$, implying, from the structure of
 $Null(H_{m+1})$, that $\nabla f(x) = A^T\lambda$ for some
$\lambda$, hence that
$x_{i+1}$ is a KT point with $\lambda$ the vector of
 Lagrange multipliers. If
 $x_{i+1}$ is not a KT point,  we can generate descent directions
by taking 
$$q = QH_{m+1}\nabla f(x)   \eqno{(19)}
$$
with $Q$ symmetric positive definite. We obtain therefore a large class
of methods with four parameter matrices ($H_1,W,V,Q$).

Some well-known methods in the literature correspond to taking 
$W=I$ in (19) and building the Abaffian as follows.

\begin{description}
\item[(a)]The {\it reduced gradient method} of Wolfe. \ \
 $H_{m+1}$ is built via the implicit LU method.
\item[(b)]The {\it projection method} of Rosen. \ \
 $H_{m+1}$ is built via the Huang
  method.
\item[(c)]The {\it  Goldfarb and Idnani method}. \ \
$H_{m+1}$ is built via a modification of Huang method where
$H_1$ is a symmetric positive definite approximation of the inverse
Hessian of $f(x)$.
\end{description}
To deal with linear inequality constraints there are two approaches in
literature.
\begin{itemize}
\item The {\it active set} method.  
 Here the set of equality constraints is augmented with some inequality constraints,
whose selection varies in the course of the process till the final set
of active constraints is determined. Adding or cancelling a single constraint
corresponds to a rank-one correction to the matrix defining the active
set. The corresponding change in the Abaffian can be performed in order two
operations, see
  Zhang (1995). 
\item The {\it standard form} approach.
Here one uses slack variables to put the problem in the following equivalent
{\it standard form}
$$min \ f(x),$$
with the constraints
$$Ax = b, \ \ x \ge 0.$$
\end{itemize}
If $x_1$ satisfies the above constraints, then a sequence of feasible
points is generated by the  iteration
$$
x_{i+1} = x_i - \alpha_i\beta_iH_{m+1}\nabla f(x)    \eqno(20)
$$
where $\alpha_i$ may be chosen by a line search along vector
   $H_{m+1}\nabla f(x)$, while
 $\beta_i > 0$ is chosen to avoid violation of the nonnegativity constraints.

If $f(x)$ is nonlinear, then $H_{m+1}$ can usually be determined once for
all at the first iteration, since generally
 $\nabla f(x)$  changes from point to point, allowing the determination
of a new search direction. However if  $f(x) = c^Tx$ is linear, in which
case we obtain the LP problem, then to get a new search direction we have
to change 
 $H_{m+1}$. We already observed in section 4 that the simplex method
in the ABS formulation is obtained by constructing the matrix
  $H_{m+1}$ via the implicit LX method and at each step modifying one
of the unit vectors used to build the Abaffian. One can show
 that the  Karmarkar method 
  corresponds to Abaffians built via a variation
 of the Huang algorithm, where the initial matrix is
  $H_1 = Diag(x_i)$ and is changed at every iteration (whether the 
  update of the Abaffian can be performed in order two operations is
  still an open question). We expect that better methods may be obtained
  by exploiting all parameters available in the ABS class.

\section{ABSPACK and its numerical performance}

The ABSPACK  project (based upon a collaboration between the University of Bergamo,
the Dalian University of Technology and the Czech Academy of Sciences)  
aims at producing a mathematical package for solving
linear and nonlinear systems and optimization problems using the ABS
algorithms. The project will take several years for completion, in view
of the substantial work needed to test  the  alternative ways
ABS methods can be implemented (via different linear algebra formulations
of the process,  different possibilities of reprojections,  different
possible block formulations etc.) and of the necessity of comparing  the
produced software with the established packages in the market (e.g. 
MATLAB, LINPACK, LAPACK,  UFO ...). It is expected that the software
will be documented in a forthcoming monograph and willl be made available
to general users.

Presently FORTRAN 77 implementations have been made
of several versions of the following ABS algorithms for  solving  linear
systems:
\begin{enumerate}
\item  The Huang  and the modified Huang algorithms in two different
       linear algebra versions of the process, for solving determined,
	underdetermined  and overdetermined systems,
       for  a solution of least Euclidean norm
\item  The implicit LU and implicit LX algorithm for determined, underdetermined
       and overdetermined linear systems, for a solution of basic type
\item  The implicit QR algorithm for determined, underdetermined and
       overdetermined linear systems, for a solution of basic type
\item  The above algorithms for some structured problems, 
       namely  KT equations  and banded matrices.
\end{enumerate}
For  a full presentation of the above methods  and their comparison with
NAG, LAPACK, LINPACK and UFO codes see Bodon,  Spedicato and Luksan(2000a,b,c).
Some results are presented at the end of this section.
There the  columns refer respectively to: 
the problem, the dimension, the  algorithm, the relative solution error
(in Euclidean norm), the relative residual error in Euclidean norm
(i.e. ratio of residual
norm over norm of right hand side), the computed rank and the time in
seconds. Computations have been performed in double precision on  a
Digital Alpha workstation with machine zero about $10^{-17}$. All test
problems have been generated with integer entries or powers of two such
that all entries are exactly represented in the machine and 
the right hand side can be computed exactly, so that the given solution
is an exact solution of the problem as it is represented in the machine.
Comparison is given with some LAPACK  codes, including those based
upon singular value decomposition ({\it svd}) and rank revealing QR
factorization  ({\it gqr}).

Analysis of all obtained results indicates:
\begin{enumerate}
\item  Modified Huang is generally the most accurate ABS algorithm and
compares in accuracy with the best LAPACK solvers based upon singular value
factorization and rank revealing QR factorization; also the estimated ranks
are usually the same.
\item  On problems where the numerical estimated rank is much less than the
dimension, one of the versions  of modified Huang is much faster than the
LAPACK codes using SVD or rank revealing QR factorization, by even 
a factor more than
 100. This is due to the fact 
that once an equation is recognized as
dependent it does not contribute to the general overhead in ABS algorithms.
\item Modified Huang is  more accurate than other
ABS methods and faster than classical methods on KT equations.
\end{enumerate}
It should be noted that the performance of the considered ABS algorithms in
term of time  could be improved by developing block versions.
ABS methods are moreover expected to be faster than many classical methods
on vector and parallel computers, see Bodon (1993).
\newpage
\begin{center}
{\bf Numerical Results}
\end{center}
\small
\begin{verbatim}   


    RESULTS ON DETERMINED  LINEAR SYSTEMS
 
    Condition  number: 0.21D+20
    IDF2    2000     huang2       0.10D+01  0.69D-11    2000   262.00
    IDF2    2000     mod.huang2   0.14D+01  0.96D-12       4     7.00
    IDF2    2000     lu lapack    0.67D+04  0.18D-11    2000    53.00
    IDF2    2000     qr lapack    0.34D+04  0.92D-12    2000   137.00
    IDF2    2000     gqr lapack   0.10D+01  0.20D-14       3   226.00
    IDF2    2000     lu linpack   0.67D+04  0.18D-11    2000   136.00

    Condition number: 0.10D+61
    IR50    1000     huang2       0.46D+00  0.33D-09    1000    36.00
    IR50    1000     mod.huang2   0.46D+00  0.27D-14     772    61.00
    IR50    1000     lu lapack    0.12D+04  0.12D+04     972     7.00
    IR50    1000     qr lapack    0.63D+02  0.17D-12    1000    17.00
    IR50    1000     gqr lapack   0.46D+00  0.42D-14     772    29.00
    IR50    1000     lu linpack   --- break-down ---                 




    RESULTS ON OVERDETERMINED SYSTEMS
 
    Condition  number: 0.16D+21
    IDF3    1050  950    huang7       0.32D+04  0.52D-13    950    31.00
    IDF3    1050  950    mod.huang7   0.14D+04  0.20D-09      2     0.00
    IDF3    1050  950    qr lapack    0.37D+13  0.83D-02    950    17.00
    IDF3    1050  950    svd lapack   0.10D+01  0.24D-14      2   145.00
    IDF3    1050  950    gqr lapack   0.10D+01  0.22D-14      2    27.00


    Condition number:  0.63D+19
    IDF3    2000  400    huang7       0.38D+04  0.35D-12    400     9.00
    IDF3    2000  400    mod.huang7   0.44D+03  0.67D-12      2     0.00
    IDF3    2000  400    impl.qr5     0.44D+03  0.62D-16      2     0.00
    IDF3    2000  400    expl.qr      0.10D+01  0.62D-03      2     0.00
    IDF3    2000  400    qr lapack    0.45D+12  0.24D-02    400     8.00
    IDF3    2000  400    svd lapack   0.10D+01  0.65D-15      2    17.00
    IDF3    2000  400    gqr lapack   0.10D+01  0.19D-14      2    12.00
\end{verbatim}    
\newpage
\begin{verbatim}
    
    
    RESULTS ON UNDERDETERMINED LINEAR SYSTEMS


    Condition number:  0.29D+18
    IDF2     400 2000    huang2       0.12D-10  0.10D-12    400    12.00
    IDF2     400 2000    mod.huang2   0.36D-08  0.61D-10      3     1.00
    IDF2     400 2000    qr lapack    0.29D+03  0.37D-14    400     9.00
    IDF2     400 2000    svd lapack   0.43D-13  0.22D-14      3    68.00
    IDF2     400 2000    gqr lapack   0.18D-13  0.24D-14      3    12.00
  
    Condition number:  0.24D+19
    IDF3     950 1050    huang2       0.00D+00  0.00D+00    950    33.00
    IDF3     950 1050    mod.huang2   0.00D+00  0.00D+00      2     1.00
    IDF3     950 1050    qr lapack    0.24D+03  0.56D-14    950    17.00
    IDF3     950 1050    svd lapack   0.17D-14  0.92D-16      2   178.00
    IDF3     950 1050    gqr lapack   0.21D-14  0.55D-15      2    26.00



    RESULTS ON KT SYSTEMS

    Condition number:  0.26D+21 
    IDF2    1000  900    mod.huang    0.55D+01  0.23D-14     16    24.00
    IDF2    1000  900    impl.lu8     0.44D+13  0.21D-03   1900    18.00
    IDF2    1000  900    impl.lu9     0.12D+15  0.80D-02   1900    21.00
    IDF2    1000  900    lu lapack    0.25D+03  0.31D-13   1900    62.00
    IDF2    1000  900    range space  0.16D+05  0.14D-11   1900    87.00
    IDF2    1000  900    null space   0.89D+03  0.15D-12   1900    93.00
 
 
    Condition number:  0.70D+20 
    IDF2    1200  600    mod.huang    0.62D+01  0.20D-14     17    36.00
    IDF2    1200  600    impl.lu8     0.22D+07  0.10D-08   1800    44.00
    IDF2    1200  600    impl.lu9     0.21D+06  0.56D-09   1800    33.00
    IDF2    1200  600    lu lapack    0.10D+03  0.79D-14   1800    47.00
    IDF2    1200  600    range space  0.11D+05  0.15D-11   1800    63.00
    IDF2    1200  600    null space   0.38D+04  0.13D-12   1800   105.00
\end{verbatim}

\end{document}